\documentclass[12pt]{article}
\usepackage{amsmath}
\usepackage{amssymb}

\numberwithin{equation}{section}

\title{
RATIONALLY-EXTENDED RADIAL OSCILLATORS AND LAGUERRE EXCEPTIONAL ORTHOGONAL POLYNOMIALS IN $k$TH-ORDER SUSYQM}
\author{C. QUESNE\\
{\small \sl Physique Nucl\'eaire Th\'eorique et Physique Math\'ematique,}\\ 
{\small \sl Universit\'e Libre de Bruxelles, Campus de la Plaine CP229,} \\ 
{\small \sl Boulevard~du Triomphe, B-1050 Brussels, Belgium} \\
{\small \sl cquesne@ulb.ac.be}}
\date{ }
\begin{document}
\baselineskip=22pt plus 1pt minus 1pt
\maketitle

\begin{abstract} 
A previous study of exactly solvable rationally-extended radial oscillator potentials and corresponding Laguerre exceptional orthogonal polynomials carried out in second-order supersymmetric quantum mechanics is extended to $k$th-order one. The polynomial appearing in the potential denominator and its degree are determined. The first-order differential relations allowing one to obtain the associated exceptional orthogonal polynomials from those arising in a ($k-1$)th-order analysis are established. Some nontrivial identities connecting products of Laguerre polynomials are derived from shape invariance.
\end{abstract}

\noindent
Running head: Laguerre Exceptional Orthogonal Polynomials


%
%
\newpage
\section{Introduction}

Since the introduction of the first families $X_1$ of Laguerre- and Jacobi-type $n$th-degree exceptional orthogonal polynomials (EOP), with $n=1$, 2, 3,~\ldots, in the context of Sturm-Liouville theory \cite{gomez09, gomez10a} and the realization of their usefulness in building new exactly solvable rational extensions of known quantum potentials \cite{cq08}, a lot of work has been devoted to generalizing these families and the associated exactly solvable potentials, as well as to providing several different (but equivalent) approaches to the problem.\par
%
%
After noting \cite{cq08} that the first extended potentials enlarged the class of known translationally shape invariant potentials in supersymmetric quantum mechanics (SUSYQM) \cite{gendenshtein, cooper, carinena}, it appeared convenient to use a SUSYQM technique to construct some additional examples of EOP and potentials \cite{bagchi09, cq09}, in agreement with previous works on algebraic deformations of shape invariant potentials \cite{gomez04a, gomez04b}.\par
%
%
Two distinct families of Laguerre and Jacobi $X_m$ EOP, now labelled as type I and type II, were constructed for arbitrary large $m$ \cite{odake09, odake10a} and their properties were thoroughly studied \cite{odake10b, ho09}. In agreement with the general definition given in Refs.\ 1 and 2, such $n$th-degree polynomials, with $n=m$, $m+1$, $m+2$,~\ldots, have the remarkable property of forming orthogonal and complete sets with respect to some positive-definite measure. They include as special cases the $X_1$ EOP first considered in Refs.\ 1, 2, 3, and 7, as well as the $X_2$ EOP introduced in Ref.\ 8.\par
%
%
In addition, the $X_m$ EOP were shown to be obtainable through several approaches, such as the Darboux-Crum transformation \cite{gomez10b, gomez11a, sasaki}, the Darboux-B\"acklund one \cite{grandati11a}, and the prepotential method \cite{ho11}.\par
%
%
Some novel procedures for generating translationally shape invariant potentials without recourse to EOP were also devised and led to examples of potentials \cite{bougie10, bougie11, ramos}, some of which were presented as new although they had already been built in Refs.\ 3 and 7.\par
%
%
Recently, the $X_m$ EOP (and associated potentials) were generalized to multi-indexed families $X_{m_1,m_2,\ldots,m_k}$, obtained through the use of multi-step Darboux algebraic transformations \cite{gomez11b}, the Crum-Adler mechanism \cite{odake11}, higher-order SUSYQM \cite{cq11} or multi-step Darboux-Ba\"cklund transformations \cite{grandati11b}.\par
%
%
In Ref.\ 25, in particular, the procedure for constructing rationally-extended radial oscillator potentials and associated Laguerre EOP in higher-order SUSYQM was considered with special emphasis on second order. The number of distinct potentials and EOP families corresponding to a given degree $\mu$ of the polynomial arising in the potential denominator (which is some definite function of $m_1$, $m_2$, \ldots, $m_k$) was determined for some small $\mu$ values and conjectured for higher ones.\par
%
%
The purpose of the present work is to extend some of the results of Ref.\ 25 to order $k>2$, to establish relations between successive EOP, and to comment on shape invariance and some of its consequences.\par
%
%
\section{Potentials and EOP in First-order SUSYQM}

To start with, let us briefly review the construction of rationally-extended radial oscillator potentials and corresponding Laguerre $X_m$ EOP in first-order SUSYQM.\par
%
%
As usual, the radial oscillator potential (in units $\hbar=2m=1$)
\begin{equation}
  V_l(x) = \frac{1}{4} \omega^2 x^2 + \frac{l(l+1)}{x^2}, 
\end{equation}
where $\omega$ and $l$ denote the oscillator frequency and the angular momentum quantum number, respectively, is defined on the half-line $0 < x < \infty$. The corresponding Schr\"odinger equation has an infinite number of bound-state wavefunctions, which, up to some normalization factor, can be written as
\begin{equation}
  \psi^{(l)}_{\nu} \propto x^{l+1} e^{- \frac{1}{4} \omega x^2} L^{(l + \frac{1}{2})}_{\nu}(\tfrac{1}{2} \omega
  x^2) \propto \eta_l(z) L^{(\alpha)}_{\nu}(z), \qquad \nu = 0, 1, 2, \ldots,
\end{equation}
with
\begin{equation}
  z = \tfrac{1}{2} \omega x^2, \qquad \alpha = l + \tfrac{1}{2}, \qquad \eta_l(z) = z^{\frac{1}{4}(2\alpha+1)}
  e^{- \frac{1}{2}z},  \label{eq:z}
\end{equation}
and $L^{(\alpha)}_{\nu}(z)$ some Laguerre polynomial \cite{gradshteyn}. The associated bound-state energies are given by $E^{(l)}_{\nu} = \omega (2\nu + l + \tfrac{3}{2}) = \omega (2\nu + \alpha + 1)$.\par
%
%
In first-order SUSYQM \cite{cooper}, one considers a pair of SUSY partners
\begin{equation}
\begin{split}
  & H^{(+)} = A^{\dagger} A = - \frac{d^2}{dx^2} + V^{(+)}(x) - E, \quad H^{(-)} = A A^{\dagger} = 
       - \frac{d^2}{dx^2} + V^{(-)}(x) - E, \\
  & A^{\dagger} = - \frac{d}{dx} + W(x), \quad A = \frac{d}{dx} + W(x), \quad V^{(\pm)}(x) = W^2(x) \mp 
       W'(x) + E,  
\end{split}
\end{equation}
which intertwine with the first-order differential operators $A$ and $A^{\dagger}$ as $A H^{(+)} = H^{(-)} A$ and $A^{\dagger} H^{(-)} = H^{(+)} A^{\dagger}$. Here $W(x)$ is the superpotential, which can be expressed as $W(x) = - \bigl(\log \phi(x)\bigr)'$ in terms of a (nodeless) seed solution $\phi(x)$ of the initial Schr\"odinger equation
\begin{equation}
  \left(- \frac{d^2}{dx^2} + V^{(+)}(x)\right) \phi(x) = E \phi(x), 
\end{equation}
$E$ is the factorization energy, assumed smaller than or equal to the ground-state energy $E^{(+)}_0$ of $V^{(+)}$, and a prime denotes a derivative with respect to $x$. Except in Sec.\ 5, we shall only consider here the case where $E < E^{(+)}_0$, in which occurrence $\phi(x)$ is nonnormalizable, and we shall assume that the same holds true for $\phi^{-1}(x)$. Then $H^{(+)}$ and $H^{(-)}$ turn out to be isospectral.\par
%
%
{}For $V_l(x)$, there are two different types of seed solutions $\phi(x)$ with such properties, namely
\begin{equation}
  \phi^{\rm I}_{lm}(x) = \chi^{\rm I}_l(z) L^{(\alpha)}_m(-z) \propto x^{l+1} e^{\frac{1}{4} \omega x^2} 
  L^{(l+\frac{1}{2})}_m(- \tfrac{1}{2} \omega x^2),
\end{equation}
\begin{equation}
  \phi^{\rm II}_{lm}(x) = \chi^{\rm II}_l(z) L^{(-\alpha)}_m(z) \propto x^{-l} e^{-\frac{1}{4} \omega x^2} 
  L^{(-l-\frac{1}{2})}_m(\tfrac{1}{2} \omega x^2),
\end{equation}
with
\begin{equation}
  \chi^{\rm I}_l(z) = z^{\frac{1}{4}(2\alpha+1)} e^{\frac{1}{2}z}, \qquad \chi^{\rm II}_l(z) = 
  z^{-\frac{1}{4}(2\alpha-1)} e^{-\frac{1}{2}z}, 
\end{equation}
and corresponding energies $E^{\rm I}_{lm} = - \omega(\alpha + 2m + 1)$, $E^{\rm II}_{lm} = - \omega(\alpha - 2m - 1)$, respectively. Note that for type II, $\alpha$ must be greater than $m$.\par
%
%
To obtain some rationally-extended radial oscillator potentials $V_{l,{\rm ext}}(x)$ with a given $l$, we have to start from a conventional radial oscillator potential $V_{l'}(x)$ with some different $l'$. We then get
\begin{equation}
\begin{split}
  & V^{(+)} = V_{l'}, \qquad V^{(-)}(x) = V_{l,{\rm ext}}(x) + C = V_l(x) + V_{l,{\rm rat}}(x) + C, \\
  & V_{l,{\rm rat}}(x) = - 2\omega \left\{\frac{\dot{g}^{(\alpha)}_m}{g^{(\alpha)}_m} + 2z 
        \left[\frac{\ddot{g}^{(\alpha)}_m}{g^{(\alpha)}_m} - \left(\frac{\dot{g}^{\alpha)}_m}
        {g^{(\alpha)}_m}\right)^2\right]\right\}, 
\end{split}  \label{eq:partners}
\end{equation}
where a dot denotes a derivative with respect to $z$ and $C$ is some constant. According to the choice made for the seed function $\phi(x)$, we may distinguish the two cases
\begin{eqnarray}
  & (i) \; &  l' = l-1, \quad \phi = \phi^{\rm I}_{l-1,m}, \quad g^{(\alpha)}_m(z) = L^{(\alpha-1)}_m(-z), \quad
         C = - \omega; \label{eq:type-I} \\
  & (ii) \; &  l' = l+1, \quad \phi = \phi^{\rm II}_{l+1,m}, \quad g^{(\alpha)}_m(z) = L^{(-\alpha-1)}_m(z), \quad
         C = \omega. \label{eq:type-II}
\end{eqnarray}
The resulting extended potential is nonsingular on the half-line for any $m=1$, 2,~\ldots (and $\alpha$ large enough for type II).\par
%
%
The bound-state wavefunctions $\psi^{(-)}_{\nu}(x)$ of $V^{(-)}$ can be obtained by acting with $A$ on those of $V^{(+)}$, $\psi^{(+)}_{\nu}(x) \propto \eta_{l'}(z) L^{(\alpha')}_{\nu}(z)$ (with $\alpha' = l' + \frac{1}{2}$), and are given by
\begin{equation}
  \psi^{(-)}_{\nu}(x) \propto \frac{\eta_l(z)}{g^{(\alpha)}_m(z)} y^{(\alpha)}_n(z), \qquad n = m + \nu, 
  \qquad \nu=0, 1, 2, \ldots,  \label{eq:partner-wf}
\end{equation}
where the $n$th-degree polynomial $y^{(\alpha)}_n(z)$ satisfies the differential equation
\begin{equation}
  \left[z \frac{d^2}{dz^2} + \left(\alpha + 1 - z - 2z \frac{\dot{g}^{(\alpha)}_m}{g^{(\alpha)}_m}\right)
  \frac{d}{dz} + (z - \alpha) \frac{\dot{g}^{(\alpha)}_m}{g^{(\alpha)}_m} + 
  z \frac{\ddot{g}^{(\alpha)}_m}{g^{(\alpha)}_m}\right] y^{(\alpha)}_n(z) = (m - n)
  y^{(\alpha)}_n(z). \label{eq:EOP-eq} 
\end{equation}
The orthonormality and completeness of $\psi^{(-)}_{\nu}(x)$, $\nu=0$, 1, 2,~\ldots, on the half-line imply that the polynomials $y^{(\alpha)}_n(z)$, $n = m+\nu$, $\nu=0$, 1, 2,~\ldots,  form an orthogonal and complete set  with respect to the positive-definite measure $z^{\alpha} e^{-z} \bigl(g^{(\alpha)}_m\bigr)^{-2} dz$. According to the choice made for $\phi(x)$, such polynomials belong to the $L1$ or $L2$ family and are denoted by $L^{\rm I}_{\alpha, m, n}(z)$ or $L^{\rm II}_{\alpha, m, n}(z)$. They are conventionally normalized in such a way that their highest-degree term is given by $(-z)^n/[(n-m)! m!]$ multiplied by $(-1)^m$ or 1, respectively \cite{odake09, odake10a, ho09}.\footnote{It is worth noting, however, that in Ref.\ 15, there is an additional factor $- (\alpha + n - 2m + 1)$ in the definition of $L^{\rm II}_{\alpha, m, n}(z)$.}\par
%
%
\section{\boldmath Potentials and EOP in $k$th-order SUSYQM}

Going from first- to $k$th-order SUSYQM \cite{andrianov, bagrov, samsonov, bagchi99, aoyama, fernandez} is achieved by replacing the first-order differential operators $A$, $A^{\dagger}$ by some $k$th-order ones $\cal A$, ${\cal A}^{\dagger}$ in such a way that the SUSY partner Hamiltonians
\begin{equation}
  H^{(1)} = - \frac{d^2}{dx^2} + V^{(1)}(x), \qquad H^{(2)} = - \frac{d^2}{dx^2} + V^{(2)}(x) 
\end{equation}
intertwine with them as ${\cal A} H^{(1)} = H^{(2)} {\cal A}$ and ${\cal A}^{\dagger} H^{(2)} = H^{(1)} {\cal A}^{\dagger}$. Here, we restrict ourselves to the reducible case, which could be formulated in the $k$th-order parasupersymmetric language as well \cite{khare}. The operator $\cal A$ can then be factorized into a product $A^{(k)} A^{(k-1)} \ldots A^{(1)}$ of first-order differential operators $A^{(i)} = \frac{d}{dx} + W^{(i)}(x)$, $i=1$, 2, \ldots, $k$ (and similarly for ${\cal A}^{\dagger}$).\par
%
%
In terms of $k$ seed functions $\phi_1, \phi_2, \ldots, \phi_k$ of the starting Hamiltonian $H^{(1)}$, the $k$ functions $W^{(i)}(x)$, $i=1$, 2, \ldots, $k$, can be expressed as $W^{(i)}(x) = - \bigl(\log \phi^{(i)}(x)\bigr)'$, where
\begin{equation}
  \phi^{(i)}(x) = A^{(i-1)} A^{(i-2)} \ldots A^{(1)} \phi_i = \frac{{\cal W}(\phi_1, \phi_2, \ldots, \phi_i)}
  {{\cal W}(\phi_1, \phi_2, \ldots, \phi_{i-1})}  \label{eq:phi-i}
\end{equation}
and ${\cal W}(\phi_1, \phi_2, \ldots, \phi_i)$ denotes the Wronskian of $\phi_1(x)$, $\phi_2(x)$, \ldots, $\phi_i(x)$. The potentials of the two partner Hamiltonians are linked by the relationship
\begin{equation}
  V^{(2)}(x) = V^{(1)}(x) - 2 \frac{d^2}{dx^2} \log {\cal W}(\phi_1, \phi_2, \ldots \phi_k). \label{eq:partner}
\end{equation}
\par
%
%
As in Sec.\ 2, we start from some conventional radial oscillator potential $V^{(1)}(x) = V_{l'}(x)$ to get as a partner an extended potential with a given $l$, $V^{(2)}(x) = V_{l,{\rm ext}}(x) + C = V_l(x) + V_{l,{\rm rat}}(x) + C$, up to some additive constant $C$. Since we have two types of seed functions at our disposal and their order is irrelevant as far as the final potential is concerned, there are altogether $k+1$ possibilities for the set of $k$ seed functions, which we may denote by ${\rm I}^q {\rm II}^{k-q}$, where $q$ runs over 0, 1, \ldots, $k$. On assuming that the first $q$ functions $\phi_i$ are of type I, we may write
\begin{equation}
  \phi_i(x) = \begin{cases}
    \chi^{\rm I}_{l'}(z) L^{(\alpha')}_{m_i}(-z) & \text{if $i=1, 2, \ldots, q$}, \\ 
    \chi^{\rm II}_{l'}(z) L^{(-\alpha')}_{m_i}(z) & \text{if $i=q+1, q+2, \ldots, k$},
  \end{cases}
\end{equation}
with $\alpha' = l' + \frac{1}{2}$, $0 < m_1 < m_2 < \cdots < m_q$, and $0 < m_{q+1} < m_{q+2} < \cdots < m_k$. For $\alpha'$ large enough, these functions are nodeless on the half-line.\par
%
%
The two pure cases I$^k$ and II$^k$, corresponding to $q=k$ and $q=0$, respectively, are very easy to deal with. For the former, on assuming $l' = l - k$ and using some simple properties of Wronskians \cite{muir}, we indeed obtain
\begin{equation}
\begin{split}
  & {\cal W}(\phi_1, \phi_2, \ldots, \phi_k) = (\omega x)^{k(k-1)/2} \bigl(\chi^{\rm I}_{l-k}\bigr)^k 
      g^{(\alpha)}_{\mu}(z), \\
  & g^{(\alpha)}_{\mu}(z) = \tilde{\cal W}\bigl(L^{(\alpha')}_{m_1}(-z), L^{(\alpha')}_{m_2}(-z), \ldots, 
      L^{(\alpha')}_{m_k}(-z)\bigr). 
\end{split}  \label{eq:pure-I}
\end{equation}
Similarly, for the latter and $l' = l + k$, we get
\begin{equation}
\begin{split}
  & {\cal W}(\phi_1, \phi_2, \ldots, \phi_k) = (\omega x)^{k(k-1)/2} \bigl(\chi^{\rm II}_{l+k}\bigr)^k 
      g^{(\alpha)}_{\mu}(z), \\
  & g^{(\alpha)}_{\mu}(z) = \tilde{\cal W}\bigl(L^{(-\alpha')}_{m_1}(z), L^{(-\alpha')}_{m_2}(z), \ldots, 
      L^{(-\alpha')}_{m_k}(z)\bigr)  \label{eq:pure-II}. 
\end{split}
\end{equation}
Here $\tilde{\cal W}$ denotes Wronskians with respect to the variable $z$, while $\mu$ is the degree of the polynomial $g^{(\alpha)}_{\mu}(z)$:
\begin{equation}
  g^{(\alpha)}_{\mu}(z) = {\cal C}^{(\alpha)}_{\mu} z^{\mu} + \text{lower order terms}.  \label{eq:EOP-norm}
\end{equation}
The values of $\mu$ and ${\cal C}^{(\alpha)}_{\mu}$ can be easily found by replacing all Laguerre polynomials by their highest-degree term, e.g. $L^{(\alpha)}_m(z)$ by $(-z)^m/m!$, in the two Wronskians and evaluating the two resulting determinants. In both cases, $\mu = \sum_{i=1}^k m_i - \frac{1}{2} k(k-1)$ and ${\cal C}^{(\alpha)}_{\mu} = (-1)^{\sigma} \Delta(m_1, m_2, \ldots, m_k)/(m_1! m_2! \ldots m_k!)$, where $\Delta(m_1, m_2, \ldots, m_k) = \prod_{i=1}^{k-1} \prod_{j=i+1}^k (m_j - m_i)$ is a Vandermonde determinant of order $k$. In case I$^k$, $\sigma = 0$, while in case II$^k$, $\sigma = m_1 + m_2 + \cdots + m_k$.\par
%
%
In the mixed cases ${\rm I}^q {\rm II}^{k-q}$, $0 < q < k$, let us assume $l' = l + k - 2q$. On taking successively into account that
\begin{equation}
  \chi^{\rm II}_{l'}(z) = \chi^{\rm I}_{l'}(z) z^{-\alpha'} e^{-z},  \label{eq:I-II}
\end{equation}
and \cite{gradshteyn}
\begin{equation}
\begin{split}
  & \frac{d}{dz} L^{(\alpha')}_m(-z) = L^{(\alpha'+1)}_{m-1}(-z), \\
  & \frac{d}{dz} \bigl(z^{-\alpha'} e^{-z} L^{(-\alpha')}_m(z)\bigr) = (m+1) z^{-\alpha'-1} e^{-z} 
      L^{(-\alpha'-1)}_{m+1}(z),  
\end{split}
\end{equation}
the Wronskian can be written as \cite{grandati11b}
\begin{eqnarray}
  && {\cal W}(\phi_1, \phi_2, \dots, \phi_k) = (\omega x)^{k(k-1)/2} \bigl(\chi^{\rm I}_{l'}\bigr)^k 
      \nonumber\\ 
  && \quad \times \tilde{\cal W}\bigl(L^{(\alpha')}_{m_1}(-z), \ldots, L^{(\alpha')}_{m_q}(-z), z^{-\alpha'} e^{-z} 
      L^{(-\alpha')}_{m_{q+1}}(z), \ldots, z^{-\alpha'} e^{-z} L^{(-\alpha')}_{m_{k}}(z)\bigr)  \nonumber\\
  && = (\omega x)^{k(k-1)/2} \bigl(\chi^{\rm I}_{l'}\bigr)^k \bigl(z^{-\alpha'-k+1}e^{-z}\bigr)^{k-q}  
      \det \Gamma^{(\alpha)}_{\mu},
\end{eqnarray}
where the matrix elements of the $k \times k$ matrix $\Gamma^{(\alpha)}_{\mu}$ are defined in terms of Pochhammer's symbol by
\begin{equation}
  (\Gamma^{(\alpha)}_{\mu})_{ij} = \begin{cases}
    L^{(\alpha' + i - 1)}_{m_j - i + 1}(-z) & \text{if $j=1, 2, \ldots, q$}, \\
    (m_j + 1)_{i - 1} z^{k-i} L^{(-\alpha' - i + 1)}_{m_j + i - 1}(z) & \text{if $j=q+1, q+2, \ldots, k$},
  \end{cases}  \label{eq:gamma}
\end{equation}
for any $i=1$, 2, \ldots, $k$.\par
%
%
To put ${\cal W}(\phi_1, \phi_2, \ldots, \phi_k)$ into a form that interpolates between those for the pure cases, given in (\ref{eq:pure-I}) and (\ref{eq:pure-II}), it is convenient to factorize it as
\begin{equation}
\begin{split}
  & {\cal W}(\phi_1, \phi_2, \ldots, \phi_k) = (\omega x)^{k(k-1)/2} z^{-q(k-q)} \bigl(\chi^{\rm I}_{l'}\bigr)^q 
      \bigl(\chi^{\rm II}_{l'}\bigr)^{k-q} g^{(\alpha)}_{\mu}(z), \\
  & g^{(\alpha)}_{\mu}(z) = z^{-(k-q)(k-q-1)} \det \Gamma^{(\alpha)}_{\mu},
\end{split}  \label{eq:mixed}
\end{equation}
by using Eq.\ (\ref{eq:I-II}) again. Determining $\mu$ and ${\cal C}^{(\alpha)}_{\mu}$, as defined in (\ref{eq:EOP-norm}), is, however, not straightforward anymore, because the replacement of the Laguerre polynomials in (\ref{eq:gamma}) by their highest-degree term leads to a matrix with $k-q$ proportional columns, whose determinant vanishes.\par
%
%
To avoid such a drawback, it is appropriate to transform $\Gamma^{(\alpha)}_{\mu}$ into another matrix $\tilde{\Gamma}^{(\alpha)}_{\mu}$ with equal determinant, for which such a problem does not occur. For any $j=1$, 2, \ldots, $k$, the matrix elements of the latter may be defined by
\begin{equation}
  (\tilde{\Gamma}^{(\alpha)}_{\mu})_{ij} = \begin{cases}
    (\Gamma^{(\alpha)}_{\mu})_{ij} & \text{if $i=1, 2, \ldots, q+1$}, \\
    \sum_{r=q+1}^i \binom{i-q-1}{i-r} (\Gamma^{(\alpha)}_{\mu})_{rj} & \text{if $i=q+2, q+3, \ldots, k$}.
  \end{cases}
\end{equation}
The summation on the right-hand side of this equation can be easily carried out by using definition (\ref{eq:gamma}) and the equations \cite{gradshteyn}
\begin{equation}
\begin{split}
  & z L^{(\alpha+1)}_m(z) + (m+1) L^{(\alpha)}_{m+1}(z) = (m + \alpha + 1) L^{(\alpha)}_m(z), \\
  & L^{(\alpha-1)}_m(z) + L^{(\alpha)}_{m-1}(z) = L^{(\alpha)}_m(z).
\end{split}
\end{equation}
It can indeed be shown by induction over $i$ running over $q+2$, $q+3$, \ldots, $k$ that
\begin{equation}
  (\tilde{\Gamma}^{(\alpha)}_{\mu})_{ij} = \begin{cases}
    L^{(\alpha' + i - 1)}_{m_j - q}(-z) & \text{if $j=1, 2, \ldots, q$}, \\
    (m_j + 1)_q (m_j - \alpha' - i + q + 2)_{i-q-1} & {} \\ 
    \quad \times z^{k-i} L^{(-\alpha' - i + 1)}_{m_j + q}(z) & \text{if $j=q+1, q+2, \ldots, k$}.
  \end{cases} 
\end{equation}
\par
%
%
{}For $g^{(\alpha)}_{\mu}(z) = z^{-(k-q)(k-q-1)} \det \tilde{\Gamma}^{(\alpha)}_{\mu}$ with Laguerre polynomials replaced by their highest-degree term, we now directly get
\begin{equation}
  \mu = \sum_{i=1}^k m_i - \frac{1}{2} q(q-1) - \frac{1}{2} (k-q)(k-q-1) + q(k-q),  \label{eq:mu}
\end{equation}
\begin{equation}
  \begin{split}
    {\cal C}^{(\alpha)}_{\mu} &= (-1)^{\sigma} \frac{\Delta(m_1, m_2, \ldots, m_q) \Delta(m_{q+1}, m_{q+2},
      \ldots, m_k)}{m_1! m_2! \ldots m_k!}, \\ 
    \sigma &= \sum_{i=q+1}^k m_i + q(k-q),  
  \end{split}  \label{eq:C}
\end{equation}
where Eq.\ (\ref{eq:mu}) agrees with a result stated without proof in Ref.\ 24. As it is clear that Eqs.\ (\ref{eq:mu}) and (\ref{eq:C}), as well as the expression for the Wronskian in (\ref{eq:mixed}), include the corresponding results for pure cases (provided $\Delta(m_1, m_2, \ldots, m_q) \equiv 1$ for $q=0$), we may treat all $q$ values simultaneously.\par
%
%
It is then straightforward to see from Eq.\ (\ref{eq:partner}) that the two partner potentials $V^{(1)}(x)$ and $V^{(2)}(x)$ assume a form similar to that of $V^{(+)}(x)$ and $V^{(-)}(x)$ in Eq.\ (\ref{eq:partners}), where we only have to substitute $\mu$ for $m$, $l+k-2q$ for $l'$, and $(k-2q) \omega$ for $C$. Provided $V^{(2)}(x)$ is nonsingular, its bound-state wavefunctions $\psi^{(2)}_{\nu}(x)$ satisfy Eqs.\ (\ref{eq:partner-wf}) and (\ref{eq:EOP-eq}) with $\mu$ substituted for $m$. We will now denote the orthogonal and complete EOP $y^{(\alpha)}_n(z)$, $n = \mu + \nu$, $\nu=0$, 1, 2,~\dots, (with respect to the positive-definite measure $z^{\alpha}e^{-z} \bigl(g^{(\alpha)}_{\mu}\bigr)^{-2} dz$) by $L^{(q,k-q)}_{\alpha, m_1, m_2, \ldots, m_k, n}(z)$, where $m_1$, $m_2$, \ldots, $m_q$ belong to type I and $m_{q+1}$, $m_{q+2}$, \ldots, $m_k$ to type II. We choose to normalize them so that their highest-degree term is $(-z)^n/[(n-\mu)! m_1! m_2! \ldots m_k!]$. This agrees with the choice made in Ref.\ 23 for $k=2$, $q=0$, but differs from the conventional one for $L^{\rm I}_{\alpha, m, n}(z)$ (see at the end of Sec.\ 2).\par
%
%
\section{Relations between Successive Laguerre EOP}

The purpose of the present Section is to derive the first-order differential relations expressing the Laguerre EOP obtained in $k$th-order SUSYQM in terms of those arising after $k-1$ steps with the use of the first $k-1$ seed functions $\phi_1$, $\phi_2$, \ldots, $\phi_{k-1}$. Since the additional seed function $\phi_k$ is of type I in the pure case I$^k$, but of type II in all the remaining cases, we have to treat them separately.\par
%
%
In the former case, where we start from $V_{l-k}(x)$, we arrive after $k-1$ steps at an extended potential $V_{l-1,{\rm ext}}(x)$, which assumes a form similar to that of $V_{l,{\rm ext}}(x)$ but with $g^{(\alpha)}_{\mu}(z)$ replaced by $g^{(\alpha-1)}_{\mu_1}(z)$, where $\mu_1 = \sum_{i=1}^{k-1} m_i - \frac{1}{2} (k-1)(k-2)$. From the bound-state wavefunctions $\eta_{l-1}(z) y^{(\alpha-1)}_{n_1}(z)/g^{(\alpha-1)}_{\mu_1}(z)$, $n_1=\mu_1$, $\mu_1 + 1$,~\ldots, of $V_{l-1,{\rm ext}}(x)$, we can obtain those of $V_{l,{\rm ext}}(x)$, $\eta_l(z) y^{(\alpha)}_n(z)/g^{(\alpha)}_{\mu}(z)$, $n=\mu$, $\mu + 1$,~\ldots, by acting with the differential operator $A^{(k)}$ defined in Sec.\ 3, namely
\begin{equation}
  A^{(k)} = \omega x \left(\frac{d}{dz} - \frac{1}{2} - \frac{2\alpha-1}{4z} - \frac{\dot{g}^{(\alpha)}_{\mu}}
  {g^{(\alpha)}_{\mu}} + \frac{\dot{g}^{(\alpha-1)}_{\mu_1}}{g^{(\alpha-1)}_{\mu_1}}\right).
\end{equation}
On using (\ref{eq:z}), we get the following equations
\begin{equation}
  {\cal O}^{(\alpha)}_{\mu_1, \mu} L^{(k-1,0)}_{\alpha-1,m_1,m_2,\ldots,m_{k-1},n_1}(z) = D
      L^{(k,0)}_{\alpha,m_1,m_2,\ldots,m_k,n}(z),  \label{eq:relation} 
\end{equation}
\begin{equation}
  {\cal O}^{(\alpha)}_{\mu_1, \mu} = \frac{1}{g^{(\alpha-1)}_{\mu_1}} \left[g^{(\alpha)}_{\mu} \left(
      \frac{d}{dz} - 1\right) - \dot{g}^{(\alpha)}_{\mu}\right],  \label{eq:O}
\end{equation}
where $n_1 = \mu_1 + \nu$, $n = \mu + \nu$, and any $\nu=0$, 1, 2,~\ldots. The constant $D$ appearing in (\ref{eq:relation}) can be derived from the normalization coefficients of the two wavefunctions and SUSYQM theory. It can alternatively be obtained by comparing the highest-degree terms on both sides of (\ref{eq:relation}). The result reads
\begin{equation}
  D = (-1)^{m_k - k} \prod_{i=1}^{k-1} (m_k - m_i).  \label{eq:D}
\end{equation}
\par
%
%
In the latter case, where we start from $V_{l+k-2q}(x)$, $0 \le q < k$, we arrive after $k-1$ steps at an extended potential $V_{l+1,{\rm ext}}(x)$, associated with a polynomial $g^{(\alpha+1)}_{\mu_1}(z)$, where $\mu_1 = \sum_{i=1}^{k-1} m_i - \frac{1}{2} q(q-1) - \frac{1}{2} (k-q-1)(k-q-2) + q(k-q-1)$. As in the previous case, from its bound-state wavefunctions $\eta_{l+1}(z) y^{(\alpha+1)}_{n_1}(z)/g^{(\alpha+1)}_{\mu_1}(z)$, $n_1=\mu_1$, $\mu_1 + 1$,~\ldots, those of $V_{l,{\rm ext}}(x)$ are obtained by application of 
\begin{equation}
  A^{(k)} = \omega x \left(\frac{d}{dz} + \frac{1}{2} + \frac{2\alpha+1}{4z} - \frac{\dot{g}^{(\alpha)}_{\mu}}
  {g^{(\alpha)}_{\mu}} + \frac{\dot{g}^{(\alpha+1)}_{\mu_1}}{g^{(\alpha+1)}_{\mu_1}}\right).
\end{equation}
The resulting equation is now
\begin{equation}
  {\cal O}^{(\alpha)}_{\mu_1, \mu} L^{(q,k-q-1)}_{\alpha+1,m_1,m_2,\ldots,m_{k-1},n_1}(z) = D
      L^{(q,k-q)}_{\alpha,m_1,m_2,\ldots,m_k,n}(z),  \label{eq:relation-bis}  
\end{equation}
with
\begin{equation}
  {\cal O}^{(\alpha)}_{\mu_1, \mu} = \frac{1}{g^{(\alpha+1)}_{\mu_1}} \left[g^{(\alpha)}_{\mu} \left( z
      \frac{d}{dz} + \alpha + 1\right) - z \dot{g}^{(\alpha)}_{\mu}\right],  \label{eq:O-bis}
\end{equation}
\begin{equation}
  D = (-1)^{k-q-1} (\alpha + \nu + k - 2q - m_k) \prod_{i=q+1}^{k-1} (m_k - m_i).  \label{eq:D-bis}
\end{equation}
\par
%
%
\section{Shape invariance and some of its consequences}

It is easy to see that if a rationally-extended radial oscillator potential $V_{l,{\rm ext}}(x)$ of the type considered in Secs.\ 3 and 4 has a ground-state wavefunction equal to $\eta_l(z) g^{(\alpha+1)}_{\mu}(z)/g^{(\alpha)}_{\mu}(z)$ up to some constant multiplicative factor, then it is (translationally) shape invariant. Such a property is based on the fact that if we consider a pair of partner potentials $(\bar{V}^{(+)}(x), \bar{V}^{(-)}(x))$ with $\bar{V}^{(+)}(x) = V_{l,{\rm ext}}(x)$ and if we take the above-mentioned wavefunction as factorization function and its eigenvalue $E^{(l)}_0$ as factorization energy, then the corresponding superpotential $\bar{W}(x)$ can be separated into $\bar{W}(x) = \bar{W}_1(x) + \bar{W}_2(x)$ with
\begin{equation}
  \bar{W}_1(x) = \frac{1}{2} \omega x - \frac{l+1}{x}, \qquad \bar{W}_2(x) = - \omega x 
  \left(\frac{\dot{g}^{(\alpha+1)}_{\mu}}{g^{(\alpha+1)}_{\mu}} - \frac{\dot{g}^{(\alpha)}_{\mu}}
  {g^{(\alpha)}_{\mu}}\right),
\end{equation}
and the partner potential is such that
\begin{equation}
  \bar{V}^{(-)}(x) - E^{(l)}_0 = \bigl[V_{l+1,{\rm ext}}(x) - E^{(l+1)}_0\bigr] + 2\omega.  \label{eq:SI}
\end{equation}
Conversely, if Eq.\ (\ref{eq:SI}) is satisfied, then the lowest-degree EOP $y^{(\alpha)}_{\mu}(z)$ is proportional to $g^{(\alpha+1)}_{\mu}(z)$.\par
%
%
Since the shape invariance of $V_{l,{\rm ext}}(x)$ has been explicitly shown in Ref.\ 26, in all cases $0 \le q \le k$, we may write
\begin{equation}
  L^{(q,k-q)}_{\alpha,m_1,m_2,\ldots,m_k,\mu}(z) = \frac{(-1)^{\sum_{i=1}^q m_i - \frac{1}{2}q(q-1) 
  - \frac{1}{2}(k-q)(k-q-1)}}{\Delta(m_1,m_2,\ldots,m_q) \Delta(m_{q+1},m_{q+2},\ldots,m_k)} 
  g^{(\alpha+1)}_{\mu}(z), 
\end{equation}
where the multiplicative factor has been found by comparing the highest-degree terms on both sides of the equation.\par
%
%
We can now combine  this result with Eqs.\ (\ref{eq:relation}), (\ref{eq:O}), (\ref{eq:D}), (\ref{eq:relation-bis}), (\ref{eq:O-bis}), and (\ref{eq:D-bis}), where we now assume $\nu=0$, $n_1=\mu_1$, $n=\mu$, to derive the two identities
\begin{equation}
  \left[g^{(\alpha)}_{\mu} \left(\frac{d}{dz} - 1\right) - \dot{g}^{(\alpha)}_{\mu}\right] g^{(\alpha)}_{\mu_1}
  = - g^{(\alpha-1)}_{\mu_1} g^{(\alpha+1)}_{\mu},
\end{equation}
\begin{equation}
  \left[g^{(\alpha)}_{\mu} \left(z \frac{d}{dz} + \alpha + 1\right) - z \dot{g}^{(\alpha)}_{\mu}\right] 
  g^{(\alpha+2)}_{\mu_1}= (\alpha + k - 2q - m_k) g^{(\alpha+1)}_{\mu_1} g^{(\alpha+1)}_{\mu},
\end{equation}
valid for $q=k$ and $0 \le q < k$, respectively. These are nontrivial $(\mu_1 + \mu)$th-degree relations connecting products of Laguerre polynomials. They can be easily checked for $k=1$ and $k=2$. In the former case, for instance, with $g^{(\alpha)}_{\mu_1} = 1$ and $g^{(\alpha)}_{\mu} = g^{(\alpha)}_m$, as given in (\ref{eq:type-I}) or (\ref{eq:type-II}), the identities reduce to $- L^{(\alpha-1)}_m(-z) - \dot{L}^{(\alpha-1)}_m(-z) = - L^{(\alpha)}_m(-z)$ and $(\alpha+1) L^{(-\alpha-1)}_m(z) - z \dot{L}^{(-\alpha-1)}_m(z) = (\alpha+1-m) L^{(-\alpha-2)}_m(z)$ for $q=1$ and $q=0$, respectively.\par
%
%
\section{Conclusion}

In the present work, we have determined the isospectral rationally-extended radial oscillator potentials resulting from conventional ones in $k$th-order SUSYQM. We have found, in particular, the polynomial $g^{(\alpha)}_{\mu}(z)$ appearing in their denominator, as well as its degree $\mu$. We have also established the first-order differential relations allowing one to obtain the corresponding Laguerre EOP from those arising in ($k-1$)th-order SUSYQM. Finally, we have proved some nontrivial identities connecting products of Laguerre polynomials, which are direct consequences of the extended potential shape invariance.\par
%
%
A similar study could be carried out for Jacobi EOP and associated rationally-extended potentials. We hope to come back to this problem in a forthcoming work.\par
%
%
\section*{Acknowledgments}

The author would like to thank Y.\ Grandati for several useful discussions.\par
%
%


\newpage
\begin{thebibliography}{99}

\bibitem{gomez09} D.\ G\'omez-Ullate, N.\ Kamran and R.\ Milson, {\em J.\ Math.\ Anal.\ Appl.} {\bf 359}, 352 (2009), arXiv:0807.3939.

\bibitem{gomez10a} D.\ G\'omez-Ullate, N.\ Kamran and R.\ Milson, {\em J.\ Approx.\ Theory} {\bf 162}, 987 (2010), arXiv:0805.3376.

\bibitem{cq08} C.\ Quesne, {\em J.\ Phys.\ A} {\bf 41}, 392001 (2008), arXiv:0807.4087.

\bibitem{gendenshtein} L.\ E.\ Gendenshtein, {\em JETP Lett.} {\bf 38}, 356 (1983).

\bibitem{cooper} F.\ Cooper, A.\ Khare and U. Sukhatme, {\em Phys.\ Rep.} {\bf 251}, 267 (1995), hep-th/9405029.

\bibitem{carinena} J.\ F.\ Cari\~ nena and A.\ Ramos, {\em J.\ Phys.\ A} {\bf 33}, 3467 (2000), hep-th/0003266.

\bibitem{bagchi09} B.\ Bagchi, C.\ Quesne and R.\ Roychoudhury, {\em Pramana J.\ Phys.} {\bf 73}, 337 (2009), arXiv:0812.1488.

\bibitem{cq09} C.\ Quesne, {\em SIGMA} {\bf 5}, 084 (2009), arXiv:0906.2331.

\bibitem{gomez04a} D.\ G\'omez-Ullate, N.\ Kamran and R.\ Milson, {\em J.\ Phys.\ A} {\bf 37}, 1789 (2004), quant-ph/0308062.

\bibitem{gomez04b} D.\ G\'omez-Ullate, N.\ Kamran and R.\ Milson, {\em J.\ Phys.\ A} {\bf 37}, 10065 (2004).

\bibitem{odake09} S.\ Odake and R.\ Sasaki, {\em Phys.\ Lett.\ B} {\bf 679}, 414 (2009), arXiv:0906.0142.

\bibitem{odake10a} S.\ Odake and R.\ Sasaki, {\em Phys.\ Lett.\ B} {\bf 684}, 173 (2010), arXiv:0911.3442.

\bibitem{odake10b} S.\ Odake and R.\ Sasaki, {\em J.\ Math.\ Phys.} {\bf 51}, 053513 (2010), arXiv:0911.1585.

\bibitem{ho09} C.-L.\ Ho, S.\ Odake and R.\ Sasaki, {\em SIGMA} {\bf 7}, 107 (2011), arXiv:0912.5447.

\bibitem{gomez10b} D.\ G\'omez-Ullate, N.\ Kamran and R.\ Milson, {\em J.\ Phys.\ A} {\bf 43}, 434016 (2010), arXiv:1002.2666.

\bibitem{gomez11a} D.\ G\'omez-Ullate, N.\ Kamran and R.\ Milson, arXiv:1101.5584.

\bibitem{sasaki} R.\ Sasaki, S.\ Tsujimoto and A.\ Zhedanov, {\em J.\ Phys.\ A} {\bf 43}, 315204 (2010), arXiv:1004.4711.

\bibitem{grandati11a} Y.\ Grandati, {\em Ann.\ Phys.\ (N.\ Y.)} {\bf 326}, 2074 (2011), arXiv:1101.0055.

\bibitem{ho11} C.-L.\ Ho, {\em Prog.\ Theor.\ Phys.} {\bf 126}, 185 (2011), arXiv:1104.3511.

\bibitem{bougie10} J.\ Bougie, A.\ Gangopadhyaya and J.\ V.\ Mallow, {\em Phys.\ Rev.\ Lett.} {\bf 105}, 210402 (2010), arXiv:1008.2035.

\bibitem{bougie11} J.\ Bougie, A.\ Gangopadhyaya and J.\ V.\ Mallow, {\em J.\ Phys.\ A} {\bf 44}, 275307 (2011), arXiv:1103.1169.

\bibitem{ramos} A.\ Ramos, {\em J.\ Phys.\ A} {\bf 44}, 342001 (2011), arXiv:1106.3732. 

\bibitem{gomez11b} D.\ G\'omez-Ullate, N.\ Kamran and R.\ Milson, arXiv:1103.5724.

\bibitem{odake11} S.\ Odake and R.\ Sasaki, {\em Phys.\ Lett.\ B} {\bf 702}, 164 (2011), arXiv:1105.0508.

\bibitem{cq11} C.\ Quesne, {\em Mod.\ Phys.\ Lett.\ A} {\bf 26}, 1843 (2011), arXiv:1106.1990.

\bibitem{grandati11b} Y.\ Grandati, arXiv:1108.4503.

\bibitem{gradshteyn} I.\ S.\ Gradshteyn and I.\ M.\ Ryzhik, Table of Integrals, Series, and Products (Academic Press, 1980). 

\bibitem{andrianov} A.\ A.\ Andrianov, M.\ V.\ Ioffe and D.\ N.\ Nishnianidze, {\em Phys.\ Lett.\ A} {\bf 201}, 103 (1995), hep-th/9404120.

\bibitem{bagrov} V.\ G.\ Bagrov and B.\ F.\ Samsonov, {\em Theor.\ Math.\ Phys.} {\bf 104}, 1051 (1995).

\bibitem{samsonov} B.\ F.\ Samsonov, {\em Mod.\ Phys.\ Lett. A} {\bf 11}, 1563 (1996), quant-ph/9611012.

\bibitem{bagchi99} B.\ Bagchi, A.\ Ganguly, D.\ Bhaumik and A.\ Mitra, {\em Mod.\ Phys.\ Lett.\ A} {\bf 14}, 27 (1999).

\bibitem{aoyama} H.\ Aoyama, M.\ Sato and T.\ Tanaka, {\em Nucl.\ Phys.\ B} {\bf 619}, 105 (2001), quant-ph/0106037.

\bibitem{fernandez} D.\ J.\ Fern\'andez C.\ and N.\ Fern\'andez-Garc\'\i a, {\em AIP Conf.\ Proc.}, Vol.\ 744, p.\ 236 (Amer.\ Inst.\ Phys., 2005), quant-ph/0502098.

\bibitem{khare} A.\ Khare, {\em J.\ Math.\ Phys.} {\bf 34}, 1277 (1993).

\bibitem{muir} T.\ Muir (revised and enlarged by W.\ H.\ Metzler), A Treatise on the Theory of Determinants (Dover, 1960).

\end {thebibliography}
\end{document}